# WET GRANULAR FLOW CONTROL THROUGH LIQUID INDUCED COHESION

Ahmed Jarray[1, 2,*], Vanessa Magnanimo[1], Stefan Luding[1]

1. Multi Scale Mechanics (MSM), University of Twente, NL-7500 AE Enschede, The Netherlands.
2. Research Center Pharmaceutical Engineering GmbH, Graz, Austria.
Email: a.jarray@utwente.nl

ABSTRACT

Liquid has a significant effect on the flow of wet granular assemblies. We explore the effects of liquid induced cohesion on the flow characteristics of wet granular materials. We propose a cohesion-scaling approach that enables invariant flow characteristics for different particles sizes in rotating drums. The strength of capillary forces between the particles is significantly reduced by making the glass beads hydrophobic via chemical silanization. Main results of rotating drum experiments are that liquid-induced cohesion decreases both the width of the flowing region and the velocity of the particles at the free surface, but increases the width of the creeping region as well as the dynamic angle of repose. Also, the local granular temperature in the flowing region decreases with an increase of the capillary force. The scaling methodology in the flow regimes considered (rolling and cascading regimes) yields invariant bed flow characteristics for different particle sizes.

# 1. Introduction

Wet granular flows are ubiquitous in nature (e.g. mud flow, debris flow and avalanches) with direct application to numerous industrial processes (e.g. coating, granulation and fertilizer production). A simple and practical geometry to study the flow of granular materials is the rotating drum. This apparatus has been extensively studied by many researchers including Henein et al. [1] who established a relationship between the flow of particles in a drum and the Froude number and categorized it into six flow modes; slipping, slumping, rolling, cascading, cataracting, and centrifuging. At low rotation speed of the drum, Rajchenbach [2] correlated the dynamic angle of repose to the rotation speed. Taberlet et al. [3] derived an equation to describe the S shape of the granular pile in a drum. Orpe et al. [4] found that the flowing layer is symmetric at low Froude number (*Fr*) and large ratios of particle size to drum radius (2*r*/*D*). Further on, Elperin and Vikhansky [5] proposed a model for describing the bed flow using a Mohr-Coulomb failure criterion. Sheng et al. [6] investigated the effect of particles surface roughness on the bed flow. At low rotation speed, below 10 rpm (*Fr*=0.08), the surface flow of the pile in a rotating drum is flat with a velocity that decreases from the bed surface down to the depth. The flow rate is usually the highest in the middle of the flowing layer where the angle of the slope is the steepest [7]. A stationary bulk region usually referred to as the creeping region can also be found below the flowing region. Komatsu et al. [8] measured the creeping region and found that the velocity rapidly decreases with depth in this region. Jain et al. [9] found that the velocity gradients and thicknesses of the flowing layer do not vary when the width of the drum is between 2 and 30 times the size of the particles.

While, most of these studies focused on the flow of dry particles, mechanisms governing particle flow in wet systems remain poorly understood. In fact, when a small amount of liquid is added to a pile of particles, pendular bridges form and the particles are attracted by capillary forces, creating complex structure and flow behavior. The work of Tegzes et al. [10] and Schubert [11] for instance showed that the capillary force strongly influences and changes the flow motion of particles. Brewster et al. [12] found that the presence of interparticle cohesion reduces the concavity of the free flowing, pushing it towards a flat or even slightly convex shape. Through experiments in a shear cell device, Chou et al. [13] found that the liquid content increases the segregation of particles in a rotating drum. Using Discrete Element Method (DEM) simulation, Liu et al [14] investigated the effect surface tension on the flow of wet particles. They showed that the maximum angle of stability of the flow in a rotating drum increases with the surface tension. Samadani and Kudrolli [15] showed that flow in wet granular systems is controlled by the number of liquid bridges. In a more recent work, Jarray et al. [16] investigated the effect of interstitial liquid on the dynamic angle of repose and showed that it is possible to scale the flow and obtain similar dynamic angle of repose for different particle sizes by modifying the surface properties of the particles. Therefore, the effects of capillary forces on wet granular flow must not be ignored. However, despite these efforts, the effect of the complex network of capillary bridges on the macroscopic properties of granular assemblies under dynamic conditions remains obscure.

The aim of this work is to explore the effect of liquid induced cohesion on the flow properties of wet granular assemblies at different particle sizes, and to establish a methodology allowing to control and eventually scale the flow of wet particles in a rotating drum. We complete our previous work [16] and we focus mainly on the macroscopic phenomena associated with the collective behaviour of particles. In their investigation of the flow of wet granular assemblies in a rotating drum, several authors including Nowak et al. [17], Xu et al. [18], Soria-Hoy et al. [19] and Tagzes et al. [10] varied the liquid content. In this work however, rather than changing the liquid content, we vary experimentally the capillary force and we take into account the variability of the contact angle. Chemical silanization is used to alter the surface properties of the glass beads allowing to obtain a wide range of capillary forces when mixed with ethanol-water mixtures. We begin by investigating and classifying the influence of the particle size, rotation speed and the capillary forces on the bed flow in dry and wet systems. We show how liquid induced cohesion affects the free flowing layer and the dynamic angle of repose. Then, we explore the effect of the capillary force on creeping and free surface flow by measuring the granular flow fields using particle image velocimetry (PIV). Flow characteristics, such as velocity and granular temperature are explored. Then, at the end of the paper, we propose a scaling approach that ensures particle flow similarity for different particle size.

## 2. Materials and methods

### 2.1. Capillary force and silanization procedure

The structure of wet granular assemblies is determined by a complex network of mechanical contacts and non-uniform liquid bridges connecting adjacent particles. Such bridges, for instance, keep a sandcastle standing [20] and determine the flow properties of wet granular materials. The dynamic behavior of wet granular materials can display a complex dependence on the amount and type of liquid present [21, 22, 23]. In this work, we will focus on the pendular state [24, 25], where liquid bridges are small and do not merge towards small separation distances or at particle contact. This results in a pairwise capillary force that displays little variations with the liquid volume [26, 27]. For two particles in contact (i.e. $d_0 \sim 0$, with $d_0$ the distance between the two particles) with the same radius (see Figure 1), the capillary force depends linearly on the contact angle $\theta$ times the surface tension $\gamma$ [28, 29].

$$F_{cap} = 2\pi\gamma r \cos\theta. \qquad \qquad Eq.\ (1)$$

Figure 1 shows the liquid bridge between two identical spheres. Here, $\theta$ is the liquid contact angle at the solid–liquid interfaces. Here, relatively small liquid bridges are considered and gravitational deformations of the menisci are neglected.

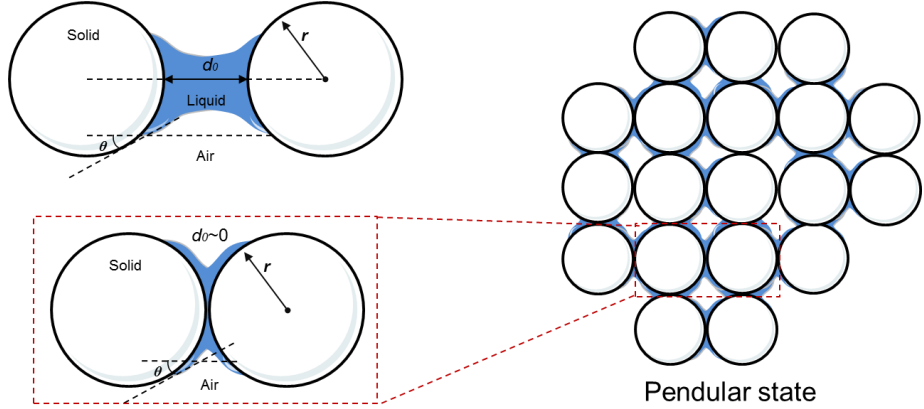

Figure 1. Pendular liquid bridge between two identical spherical particles.

In order to characterize the capillary forces generated by liquid bridges and their effect on the flow of granular assemblies, we consider separately the roles of particle size, contact angle, and surface tension. Similarly to the work of Raux et al. [30], we treat the surface of the glass beads via silanization to increase their hydrophobicity (i.e. we increase the contact angle $\theta$). Using mixtures of ethanol-water as interstitial liquid and glass beads with different hydrophobicity, the capillary force between two adjacent beads can be manipulated over a wide range of values.

Silanization is based on the adsorption, self-assembly and covalent binding of silane molecules onto the surface of glass beads. Chemical compounds used for silanization are: silanization solution I~5% ($V/V$) (5% in volume of Dimethyldichlorosilane in Heptane, Selectophore), Hydrochloric acid (HCl, 0.1 mol), Acetone and Ethanol.

The procedure for increasing the beads hydrophobicity is as follows: first, glass beads, initially hydrophilic, are cleaned for at least one hour by immersion into freshly prepared HCL solution under agitation using a rotor-stator homogenizer. Then, they are rinsed thoroughly with deionized water and oven dried for 3 hours at 60°C. Afterward, the freshly cleaned glass beads are immersed in the silanisation solution under low agitation at room temperature for one hour. The inorganic functionality of the silane reacts with the different OH groups obtained after cleaning with HCL and forms Si-OH groups that make the glass beads hydrophobic while their topography is conserved. Finally, the glass beads are rinsed with acetone and allowed to air-dry under a fume hood for 24 hours.

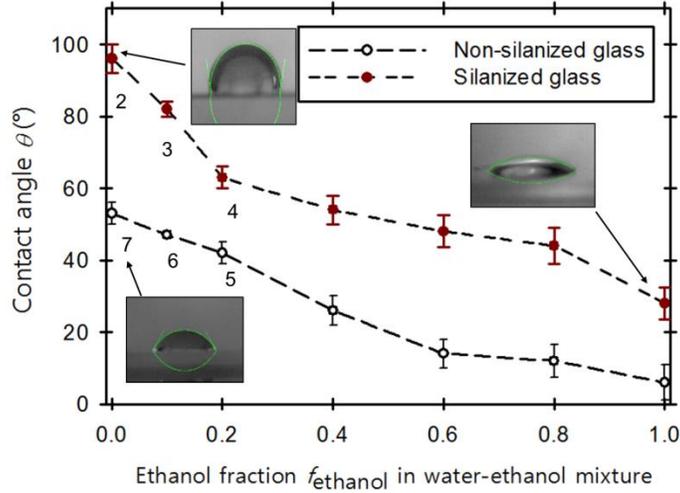

Figure 2. Contact angles of water-ethanol mixtures as a function of the ethanol fraction on silanized and non-silanized glass. Adapted from Ref. [16].

To measure the contact angle, $\theta$, liquid droplets are placed on a glass surface. Then, pictures of the droplet are taken using a MotionBLITZ EoSens camera with close-up lenses. The contact angle is then deduced by image analysis using LBADSA plugin in the open source imageJ software. The LBADSA plugin is based on the fitting of the Young-Laplace equation to the image data [31]. As shown in Figure 2, water−ethanol mixtures with different ethanol fractions give intermediate contact angles, which continuously decrease from 96° to 30° for silanized beads as the ethanol fraction in the liquid increases: the higher the contact angle, the lower the wettability of the more hydrophobic glass surface.

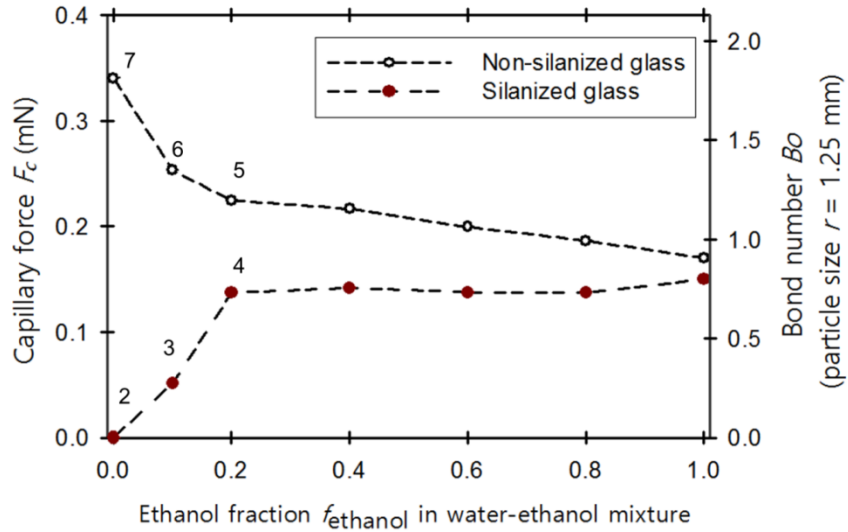

Figure 3. Capillary forces $F_c$ of water-ethanol mixtures as a function of the ethanol fraction on silanized and non-silanized glass beads. $F_c$ is measured using Equation (1).

In Figure 3, we plot the capillary force $F_{cap}$ defined in Equation (1) against the volume fraction of ethanol in the water-ethanol mixture. Surface tension values are taken from Ref. [32]. Since the contact angle $\theta$ affects the strength of the capillary force, here, differently from e.g. [33, 34], the variability of the contact angle is taken into account. The bond number $Bo$ (Equation 2), representing the ration of the capillary force to the gravitational force, is also shown along the y-axis as a function the water-ethanol fraction.

$$Bo = \frac{3\gamma \cos\theta}{2r^2 \rho_p g}, \qquad\qquad Eq.~(2)$$

where $\rho_p$ is the density of the particle.

The capillary force decreases with increasing ethanol concentration for particles with unmodified glass surface, while $F_{cap}$ increases in the case of silanized glass beads, in agreement with the literature [35]. This allows controlling the strength of the capillary force between the glass beads in the rotating drum. Overall, 1 dry and 6 different wet experiments were performed in the rotating drum for different particle sizes (0.85, 1.25 and 2 mm in radius). The wet cases correspond to the 6 left points in Figure 2. Details of the experiments are given in Table 1.

Table 1. Experimental cases used throughout this work. Wet cases correspond to the 6 left points in Figure 2.

| Cases  | Wet or dry | Silanized | Ethanol fraction | Capillary forces $F_{cap}$ (mN, for particle size $r = 1.25$ mm) | Bond number $Bo$ (for particle size $r = 1.25$ mm) |
|--------|------------|-----------|------------------|------------------------------------------------------------------|----------------------------------------------------|
| Case 1 | Dry        | no        | -                | -                                                                | -                                                  |
| Case 2 | Wet        | yes       | 0                | 0                                                                | 0                                                  |
| Case 3 | Wet        | yes       | 0.1              | 0.0517                                                           | 0.2755                                             |
| Case 4 | Wet        | yes       | 0.2              | 0.1372                                                           | 0.7315                                             |
| Case 5 | Wet        | no        | 0.2              | 0.2247                                                           | 1.1974                                             |
| Case 6 | Wet        | no        | 0.1              | 0.2533                                                           | 1.3500                                             |
| Case 7 | Wet        | no        | 0                | 0.3403                                                           | 1.8134                                             |

## 2.2 The drum apparatus

Samples of silanized and non-silanized glass particles, as described above in Table 1, are placed in a rotating drum (see Figure 4). According to Jain et al. [36], the width of the drum in the $z$ direction should be larger than 6.4×$r$, with $r$ the average radius of the particles, to neglect the effect of the wall friction on the flow characteristics. Consequently, the velocity field can be assumed to be two-dimensional. Our drum is made by a cylinder of 121 mm inner diameter, 22 mm width, larger than 6.4×$r$ of all the particles used, and held between two circular plexiglass (PMMA) plates of 5 mm thickness to allow optical access (frontal view in Figure 4 (a)). The drum is placed on a horizontal rotating axis driven by a variable-speed motor, aligned with the $z$ axis. The PMMA walls of the drum are coated with Fluorinated Ethylene Propylene (FEP) coating manufactured by CS Hyde Company to prevent wet glass beads from sticking on the wall. Images of the rotating drum are recorded using a MotionBLITZ EoSens high speed camera working at a speed of 460 fps. The angle between the top surface of the rolling bed and the horizontal plane is called the dynamic angle of repose $\theta r$ (Figure 4 (b)). We refer to the angle at the bottom of the drum as the lower dynamic angle of repose $\theta s$. The axis $y$ is normal to the flow, and we refer to the zone in the middle of the bed along the $y$ direction as the median region (See Figure 4 (b)).

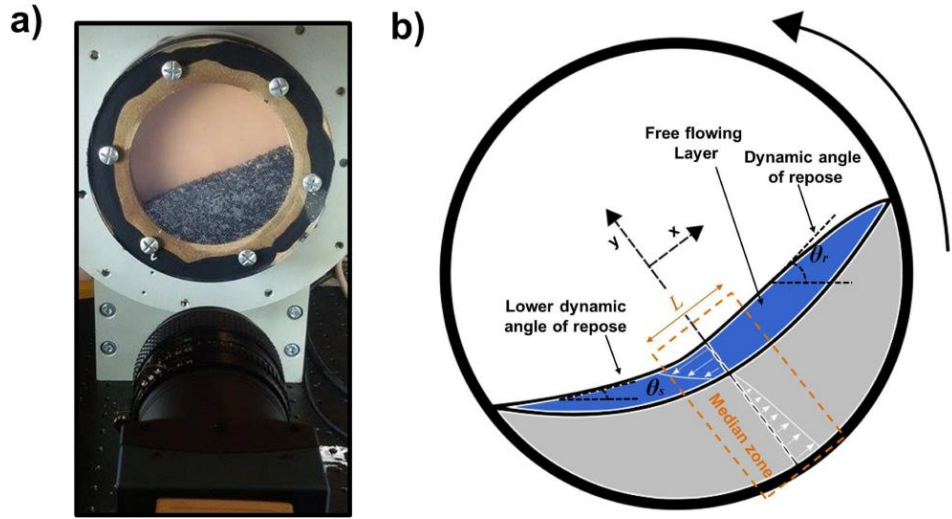

Figure 4. a) Rotating drum apparatus (back) and camera (front), b) Schematic representation of the particle flow in a rotating drum.

After each experiment, the drum is dried in an oven for one hour to let the remaining liquid from the previous experiment evaporate. Experiments were restricted to the slumping, rolling and cascading regimes in the range 0.05 < $Fr$ < 0.4, where $Fr$ is the Froude number representing the ratio of centrifugal to gravitational acceleration [37].

$$F_r = \sqrt{\frac{\omega^2 D}{2g}},  \quad\quad Eq.\ (3)$$

where $\omega$ is the angular speed of the drum, $D$ its diameter and $g$ the acceleration due to gravity. Experiments are conducted using a selected set of monodisperse glass beads of density 2500 kg/m³ with Young's modulus $Y = 6.4 \times 10^{10}$ N/m² and Poisson ratio $\nu = 0.2$. Parameters and characteristics of the drum and the glass beads are summarized in Table 2.

Throughout this work, a reference amount of liquid $V_{liq}$ is mixed with the same mass of particles (i.e. 125 g of particles) in the rotating drum. Using $V_{liq}$ = 4 ml as interstitial liquid for all the particle sizes used (i.e. 0.85, 125 and 2 mm), the granule saturation, $s$, representing the fraction of liquid in a packed bed, defined in Equation (4) is below 25%, validating that we are in the pendular state [38, 39].

$$s = \frac{V_{liq}(1-\xi)}{V_{bed}\xi}, \qquad Eq.\ (4)$$

where $\xi$ is the porosity of the bed in the rotating drum, $V_{bed}$ is the volume of the bed of particles.

Table 2. Properties of the drum and the glass beads.

| Properties | Value |
|---|---|
| Drum, $D \times L$ (mm) | 121×22 |
| Glass beads radii $r$ (mm) | 0.85, 1.25 and 2 |
| Rotation speed $\omega$ (rpm) | 5 to 45 |
| Particle density $\varrho_p$, (kg/m³) | 2500 |
| Filling level $\beta$ | 35% (~125 g) |
| Young's modulus $Y$ (N/m²) | $6.4 \times 10^{10}$ |
| Poisson ratio $\nu$ | 0.2 |
| Volumetric liquid content $V_{liq}$ (ml) | 4 |

## 2.3 Image post-processing

The images, acquired using the high-speed camera, are post-processed using the particle tracking package Trackmate within the Fiji ImageJ distribution [40]. In Figure 5, we show the images post-processing steps that we used in this work to compute the dynamic angle of repose and the velocity gradient. First, we removed the background from each image and adjusted the lightning in order for the moving particles to become markedly brighter than the background. Then, Trackmate package is used to detect the position of the particles in every frame. Particles outlines (i.e. spots) that stand out from the background were segmented and identified based on the difference of Gaussians approach [41] with an estimated particle diameter of 8 pixels. Detected particles are represented as spots or spheres with an initially constant radius. The detected spots are then imported in ParaView software [42] and converted into a sequence of tables. Then, the dynamic angle of repose is numerically computed by linear regression of the positions of the particles using an in-house python code filter. The python code detects the particles which are on the surface of the flow using a concave-hull algorithm [43].

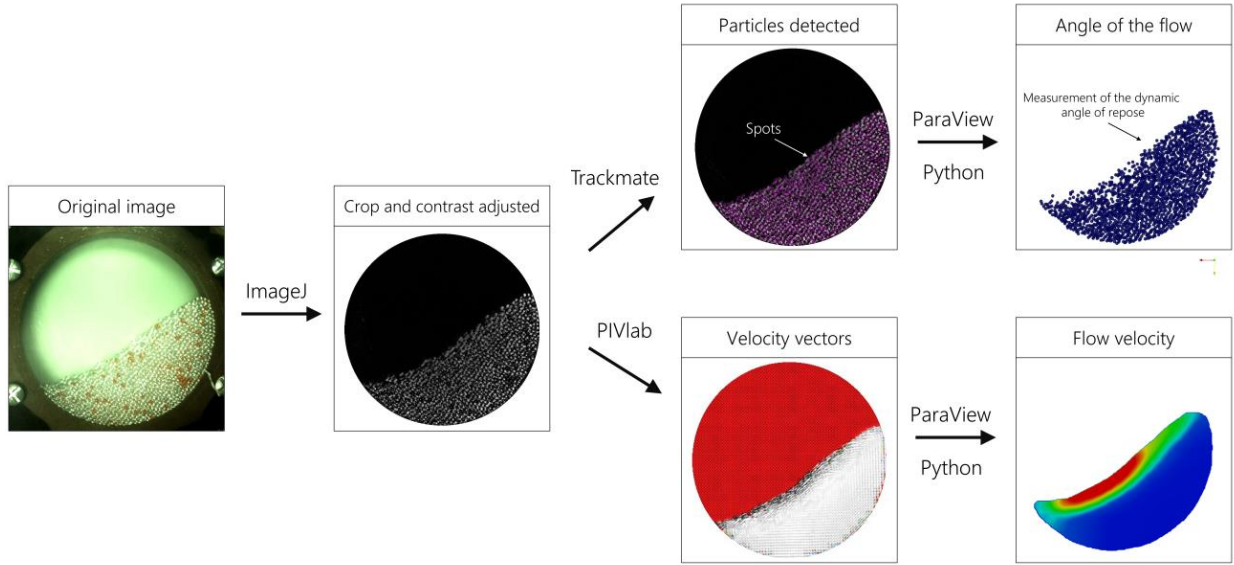

Figure 5. Post processing steps of the images acquired from the high-speed camera.

Particle Image Velocimetry (PIV) in the PIVlab package [45] in Matlab is used to measure the particles displacements and vector fields. PIV is a non-intrusive, image-based measurement technique where a deformation pattern is detected by comparing two consecutive images. First, the drum which represents the area of interest (AOI) is cut out of the digital image and divided into small sub-areas called interrogation cells (see Figure 5). Then, every two successive frames are combined into a single new frame. To obtain the velocity field between the interrogation areas in the first and second image, a multi-pass correlation algorithm with a final interrogation size of 8 pixels is used, which is of the order of the particle size. An interrogation size around the order of the particle size has also been adopted by previous authors [46, 47] to ensure smooth and reliable results. For each experimental run, we applied PIV analysis for over 120 consecutive images (i.e. over the period of 0.33 s). The final time-averaged velocity field is obtained by averaging all velocity fields that are calculated using each newly combined image. The data obtained from PIV are then imported in ParaView for further post-processing. Using several python scripts the granular temperature and the flow velocity are computed and visualized in ParaView (see section 2.4).

## 2.4 Coarse-graining and properties computation

The velocity data obtained from PIV are converted into a smooth continuous velocity field using coarse-graining [48]. For the coarse-graining of the velocity data, we follow the approach described by Weinhart et al. [48, 49]. In this approach, a cut-off Gaussian coarse-graining function is used:

$$W(x) = \vartheta^{-1}\exp(-\frac{x^2}{2\alpha^2}), \quad \text{if} \quad x := |x| < c, \quad 0 \quad \text{else,} \qquad Eq.\ (6)$$

with the coarse-graining width $\alpha$, the normalisation constant $\vartheta$, and the cutoff $c = 6r$. The width $\alpha$ is chosen equal to the particle diameter as it was advised by Weinhart et al. [48, 49].

The coarse-grained velocity field $v\ (x)$ is then given by:

$$v(x) = \frac{\sum_{i=1}^{N} m_i v_i W(x - x_i)}{\rho(x)}, \qquad Eq.\ (7)$$

where $v_i$ is the velocity of particle $i$, and $\rho(x)$ the macroscopic mass density field:

$$\rho(x) = \sum_{i=1}^{N} m_i W(x - x_i). \qquad Eq.\ (8)$$

From the velocity field we characterize the flow of particles by measuring the velocity of the particles in the drum minus the angular velocity $v_\Omega$:

$$\varphi = \sqrt{(\vec{v}_x - \vec{v}_{\Omega x})^2 + (\vec{v}_y - \vec{v}_{\Omega y})^2}. \qquad Eq.\ (5)$$

We will refer to $\varphi$ as the flow velocity.

The granular temperature $T$ of particles quantifies the inter-particle random motions caused by continuous collisions between particles or between particles and boundaries. Measurements of $T$ are useful to elucidate the variable dynamic nature of granular material flows. Since the granular temperature is analogous to the thermodynamic temperature, it should be expected that $T$ is the highest in the regions where the material exhibits fluid-like flow and lowest in those regions where the behaviour is solid-like [50]. As suggested by Ahn et al. [51] and Bonamy et al. [52], we extract the granular temperature $T$ as the variance of the velocities for a short duration of the flow (0.33 s) by subdividing the drum into elementary square cells $\sum(x, y)$ of size set equal to twice the particle diameter. We then define the velocity fluctuation of a particle $i$ as:

$$\delta\varphi_i = \sqrt{(\vec{\varphi}_{i,x} - \vec{\varphi}_{av,x}(x,y))^2 + (\vec{\varphi}_{i,y} - \vec{\varphi}_{av,y}(x,y))^2}, \qquad Eq.\ (9)$$

where $\varphi_{av}$ is the mean flow velocity value of the cell $\sum(x, y)$ that contains the particle $i$. From the velocity fluctuation we can compute the granular temperature $T_i$ using:

$$T_i = \frac{1}{2}\delta\varphi_i^2. \qquad Eq.\ (10)$$

# 3. Results and discussion

We first examine the effect of particle size and drum rotation speed on the flow characteristics in the dry case using monosized spherical particles. This is followed by results showing the effect of capillary forces on the flow. The obtained data are afterward used to establish scaling relationships.

## 3.1 Dry case: effect of the particle size and rotation speed on the flow

In a rotating drum, particles are continuously lifted to the upper part of the drum and the angle of the slope increases until the maximum angle of stability $\theta m$ (also called avalanching angle) is reached. Once this angle is exceeded, avalanche occurs and the particles slump or roll down the slope of the flowing layer to a new inferior dynamic angle of repose $\theta r$ (see Figure 4 (b)). Maximum angle of stability $\theta m$ as a function of the Froude number $Fr$, for the dry case, for particle size $r = 1.25$ mm is shown in Appendix A. Particles are then introduced again inside the lower part of the bed. There are two important criteria of the avalanche phenomenon, first, it occurs right after the slope of the free surface exceeds a threshold, and a fluid-like flow happens for a period of time equal to the avalanche duration, until the lowest energy state is reached (i.e. particles are at rest). The second criterion is that after few rotations of the drum, a continuous flow regime is reached where $\theta m \approx \theta r$.

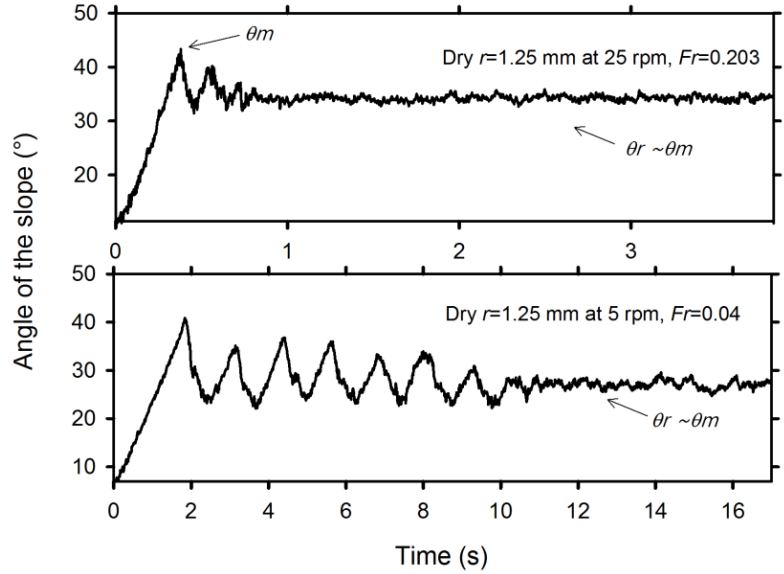

Figure 6. Transition from avalanching to continuous regime. Glass beads ($r$ = 1.25 mm) in a rotating drum.

As an example, we show in Figure 6 the dynamic angle of the flow as a function of time for dry particles of radius 1.25 mm and two rotation rates; $\omega$ = 5 rpm and $\omega$ = 25 rpm. The drum starts to rotate slowly such that the slope of the interface increases linearly. Then, successive avalanches occur with a stick-slip motion depending on the rotation speed of the drum until a steady continuous flow with small variations is obtained. A complete characterization of the avalanche dynamics of granular media in a rotating drum was carried out by Tegzes et al. [10] and similar behavior of the flow was observed.

Next, we focus on steady continuous flows and we characterize the flow in the drum using the dynamic angle of repose $\theta r$ and the flow velocity profile of the granular assemblies.

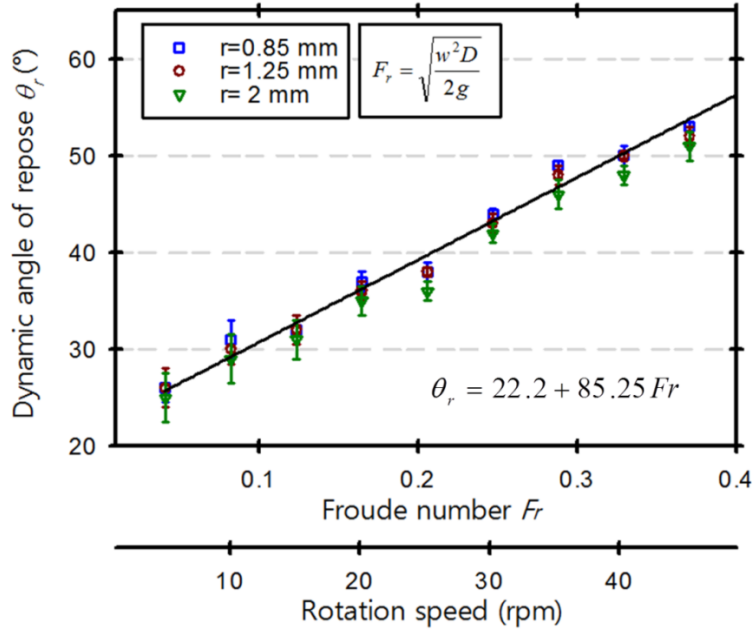

Figure 7. Dynamic angle of repose $\theta r$ as a function of the rotation speed and the Froude number $Fr$, for the dry case.

Figure 7 shows the dynamic angle of repose versus the rotation speed for the different particle sizes in the dry case (0.85, 1.25 and 2mm, case. 1). The dynamic angle of repose increases as the rotation speed increases but the effect of particle size on the dynamic angle of repose is small. This is because the particle size only has a negligible effect on the Froude number. Nevertheless, we notice that increasing the particle size slightly decreases the angle of repose at the same rotation speed.

At low rotation speed (i.e. low Froude number), the error bars of the dynamic angle of repose are large, indicating relatively high variations of the slope of the bed, reminiscent to the slumping intermittent avalanching regime. Around 15 rpm (i.e. $Fr = 0.125$), the error bars of the angle of repose become less large due the shrinking of the periodicity of the small avalanches occurring in the continuous flow. At around 20 rpm (i.e. $Fr = 0.16$), the S shape starts to form, and the lower dynamic angle of repose decreases with the formation of a curved free surface. At 30 rpm, the flowing layer acquires an even more curved shape, and displays the distinct signatures of the cascading regime. Above 45 rpm (i.e. $Fr = 0.37$), the flow becomes slightly discontinuous marking the transition from the cascading to the cataracting regime. For all particle sizes, the dynamic angle of repose collapses into one linear profile when scaling the rotation speed using the Froude number $Fr$ (see Figure 7).

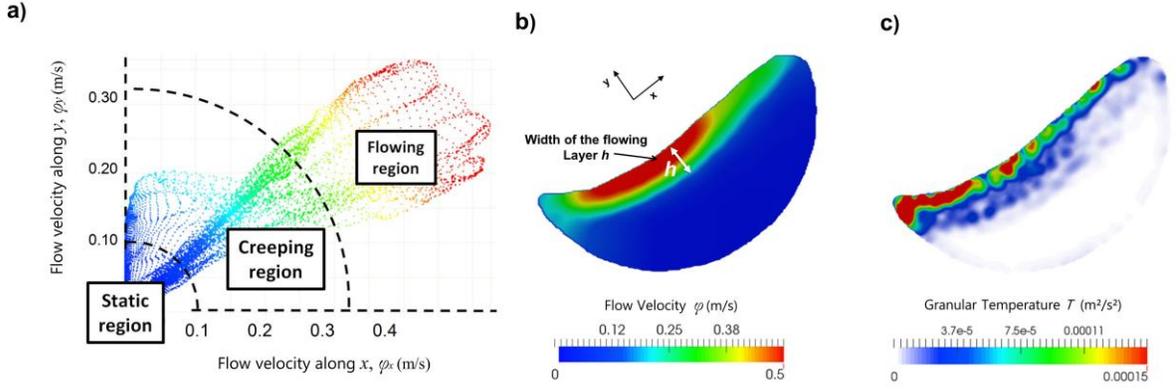

Figure 8. Snapshots of the a) vertical and horizontal velocity components, (b) flow velocity $\varphi$ and (c) granular temperature $T$ of the dry bed of 1.25 mm particles at 25 rpm ($Fr$ = 0.203).

In Figure 8 (a), we show the flow velocity components $\varphi_x$ and $\varphi_y$ of 1.25 mm particles after 10 seconds of drum rotation and at 25 rpm ($Fr$ = 0.203, in the cascading regime), where the continuous flow is reached. The flow velocity components show a jellyfish shape where the highest flow velocity magnitude is observed at the top-right tail, and the lowest is observed at the head of the jellyfish at the bottom-left. Three regions can be distinguished: 1) A flowing region represented by the tail of the jellyfish exhibiting high flow velocity comparing to the core of the bed. The maximum flow velocity in this region occurs approximately in the centre of the flowing layer. The bottom part of the drum in Figure 8 (a) represents the static zone where particles are quasi-immobile. Finally, a creeping region where the magnitude of the flow velocity $\varphi$ increases rapidly with the normal to the flow $y$. We define the boundaries of this region between the interfaces of the static region and the flowing region (i.e. where the velocity magnitude is approximately between 0.1 and 0.25 m/s). In Appendix B, we show also the velocity components $v_x$ and $v_y$ of the flow.

Figure 8 (b) shows a coarse grained field of the flow velocity $\varphi$ defined in section 3.a and averaged over 120 frames (i.e. 0.33 s). In the upper part of the bed, the intensity of the $\varphi$ is the largest, and decreases gradually along the creeping zone. However, in the lower part of the bed, the velocity flow $\varphi$ is almost equal to zero in the core and near the wall of the bed. This is because the angular velocity is subtracted from the velocity.

Figure 8 (c) shows a coarse grained field of the granular temperature. Within the core of the bed, the granular temperature is close to zero due to the absence of local velocity fluctuations. This is reminiscent of the solid-like flow where particles move collectively as a solid in a rigid body rotation. However, large fluctuations are observed within the flowing region, with several hot local zones distributed along the upper part of the flowing layer, i.e. near the free surface, which is consistent with the couette flow modelling results of Zhang & Campbell [50]. This indicates high vibrations of the particles due to binary random collisions. In a study of the flow of dry granular material in rotating drums, Chou et al. [53] also found that the granular temperature prevails only in the flowing layer in the rotating drum. This also supports the finding of Bonamy et al. [54], where they demonstrated the presence of clusters in the free flowing layer emitted by the static phase at the bottom of the drum.

## 3.2 Wet case: Effect of capillary forces

Let us first examine the dynamic angle of repose in the simplest case of non-silanized particles with pure-water as interstitial fluid. This will help us to gain knowledge on the interplay between the effect of the particle size and rotation speed when the particles are wet. The relative velocity in our experiments never exceeds 10 m/s, and since the viscosity of water and ethanol are very low (0.89 mPa.s for water and 1.074 mPa.s at 25 °C), the capillary number (ratio of the viscous force to the capillary force) $Ca$ is always below unity (see Equation 9). Hence, the capillary forces are more dominant than the viscous forces and therefore can be neglected.

$$Ca = \frac{v_r \mu}{\gamma \cos\theta} \ll 1, \qquad Eq.\ (9)$$

with $v_r$ the particle-particle relative velocity and $\mu$ the interstitial liquid viscosity.

Figure 9 shows the dynamic angle of repose versus rotation speed for three particle sizes 0.8, 1.25 and 2 mm, in wet system with pure water as interstitial liquid. Clearly, the interparticle liquid cohesion has a significant effect on the bed flow motion. As the rotation speed increases, the dynamic angle of repose increases. Smaller glass beads have higher dynamic angle of repose. This is because a decrease of the particle size decreases the capillary force, but increases the Bond number significantly more. We observe two regimes; when the rotation speed of the drum is below 35 rpm, the dynamic angle of repose increases linearly with the rotation speed. Above 35 rpm, the capillary force becomes less effective for the case of 2 mm particle size comparing to 0.85 and 1.25 mm, with a significant difference in the angles of repose over the three particle sizes.

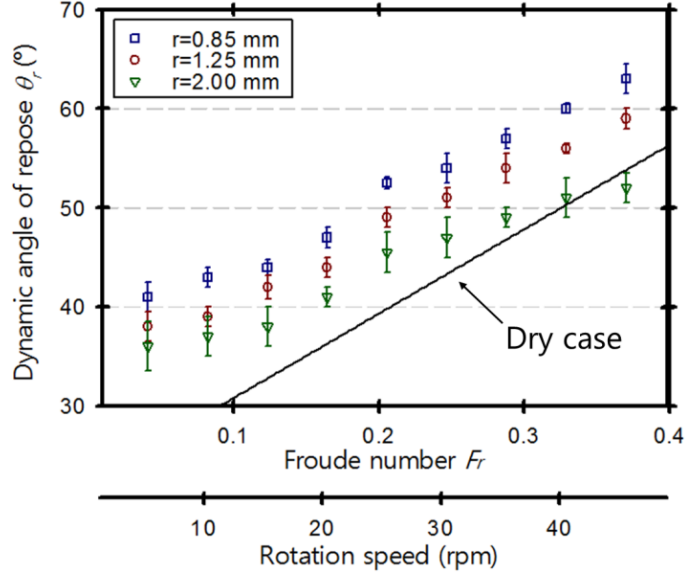

Figure 9. Dynamic angle of repose as a function of the rotation speed for different particle sizes in the wet case.

We now focus our attention on the effect of the capillary forces on the dynamic angle of repose, flow velocity gradient, and granular temperature for only one rotation speed of the drum set at 25 rpm ($Fr = 0.203$) and for particles of radius $r = 1.25$ mm. We tune the capillary force by mixing hydrophobic or hydrophilic glass beads with different ethanol-water mixtures as described in the previous section.

The dynamic angle of repose for the wet 1.25 mm particles as a function of the capillary force is shown in Figure 10. Again, we find that as the capillary force increases, the dynamic angle of repose increases. The angle of the slope of case 7 (i.e. wet, non silanized with 0% ethanol) is about 49°, larger than that of dry particles (around 38° in Figure 7). This is not surprising. As we increase the capillary force at a given rotation rate, the particles experience a greater capillary cohesion and they are dragged more to the upper part of the drum, causing an increase in the slope of the bed. Similar findings were obtained by Nowak et al. [17] and Soria-Hoyo et al. [19] but only for humid glass beads without varying the capillary forces. They showed that the angle of the flow decreases with an increase of the particle size for wet particles. Similarly, Liu et al [14] found that an increase of the surface tension of the liquid increases the angle of the slope as well as the collision frequency. Furthermore, as expected, silanized glass beads mixed with water show low dynamic angle of repose (case 2) close to that of the dry case. This is because the capillary force becomes weaker in the case of silanized surfaces due to the higher glass-liquid contact angle.

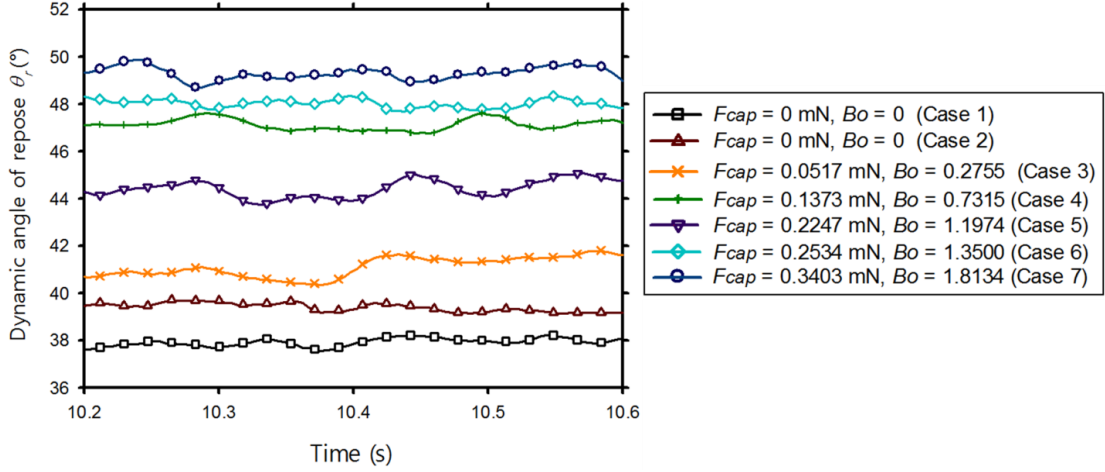

Figure 10. Smoothed dynamic angle of repose as a function of time in the continuous flow regime for different capillary forces. Particles size $r$ = 1.25 mm and drum rotation speed $\omega$ = 25 rpm ($Fr$ =0.203).

For a complete understanding of the flow of wet granular assemblies, we show in Figure 11 the gradient of the flow velocity $\varphi$, granular temperature, and velocity components of the granular flow for particle radius 1.25 mm at 25 rpm and for different capillary forces.

In the flow velocity $\varphi$ gradients profile of the drum at the top row of Figure 11, the flowing layer is easily distinguishable from the bulk. The slope of the free surface increases with increasing the capillary forces which confirms the results obtained from the measurement of the dynamic angle of repose. Whereas the flow pattern has a convex shape for dry and low cohesive particles, an almost flat shape is observed for highly cohesive particles, particularly for 100% ($V/V$) water case (case. 7 in Table 1). By visual inspection of $\varphi$ gradients in Figure 11, we can see that the flow is confined to a thin layer near the free surface. This layer is composed of the flowing region and the creeping region whose width increases with the capillary forces, and starts to acquire a symmetric profile, suggesting a gradual transition from shear flow to plug flow. At the same time, the surface flow velocity seems to decrease with increasing the capillary force. This is despite the fact that the capillary force causes a larger dynamic angle of repose, resulting in a greater kinetic energy from the increase of the down-slope direction of the gravitational acceleration. This decrease of the flow velocity can be attributed to two simultaneous effects; one effect is that the sliding of particles on the free surface is reduced by the cohesion forces exerted by the particles below them in the creeping region. The second effect is that the kinetic energy of the particles is also dissipated by the increasingly strong cohesive forces between the wet particles. This hints on the existence of a competition between inertial and capillary forces that governs the flow of wet particles.

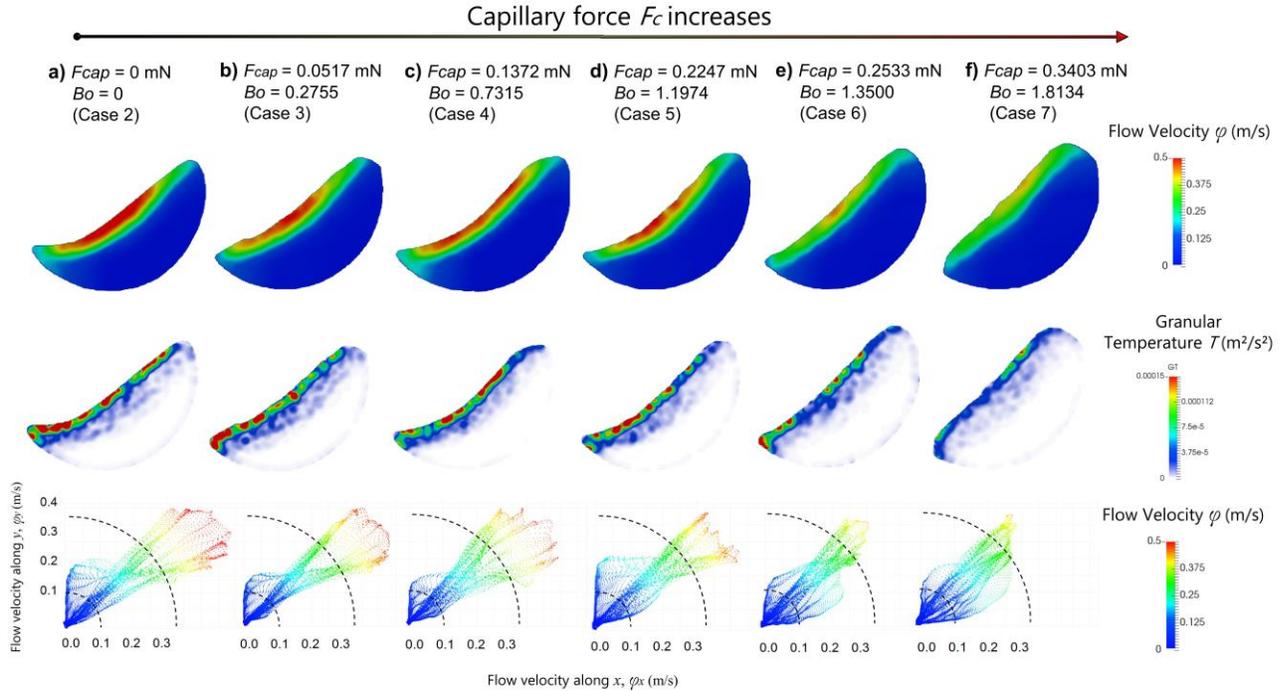

Figure 11. Snapshots of the flow velocity gradient $\varphi$ (top), granular temperature $T$ (second row) and the velocity vertical and horizontal components of the bed in the rotating drum for different capillary forces. Particles size $r = 1.25$ mm and drum rotation speed $\omega = 25$ rpm.

The values of the granular temperatures in Figure 11, second row, range from zero near the static zone, up to 0.0002 m²/s² in some spots scattered in the flowing region. The overall average low fluctuations of the flow velocity are due to a combination of capillary forces that reduces the collision frequency by cohesion, and to a relatively slow flowing regime (i.e. cascading regime). Even though silanized wet glass beads (Case. 2) flow almost like dry particles (Figure 11 (a)), we notice a decrease of the number of hot spots of granular temperature comparing to the dry case (see Figure 8 (c)), especially in the left side of the flowing region, but, nevertheless, granular flow is still dominated by the inertial forces of the individual particles in the case of silanized beads mixed with water (case 2). This is because, at this point, the capillary forces are still weak and the particles can freely roll and move randomly along the free surface. Then, as the capillary force increases, a gradual reduction of the granular temperature in the flowing and creeping regions is observed, indicating less frequent collisions between the particles especially for wet non-silanized cases (Figure 11 (e) and (f), cases 6 and 7), where the capillary forces are the highest. Particles become closely packed and act as clusters rather than individually and flow on the free surface as a bulk. This can also be seen in the flow velocity components graph at the bottom of Figure 11, showing the velocity components of the bed. As the capillary force increases, the jellyfish tail shrinks progressively, which is translated by less scattered distribution of the $x$ and $y$ components of the flow velocity in the flowing region, with particles moving in an orderly fashion on the surface of the bed.

Flow velocity $\varphi$ averaged along the normal to the flow is also illuminating. To obtain its profile, the flow velocity of the particles whose positions are between $y$ and $y+dy$ of the median zone of width $L$ (i.e. particles inside the cubic cell $(y+dy)\times L$) (see Figure 4 (b)) are averaged and then projected on a line along the $y$ direction:

$$\varphi(y) = \frac{\sum_{i=1}^{N_y} \varphi_i(y)}{N_y}, \qquad Eq.\ (10)$$

where $N_y$ is the total number of particles in the cell $(y+dy)\times L$, with $L$ the width of the median zone equals to $4\times r$ here. We performed an averaging over the width $L$ to obtain smoother representative curves with minimum noise.

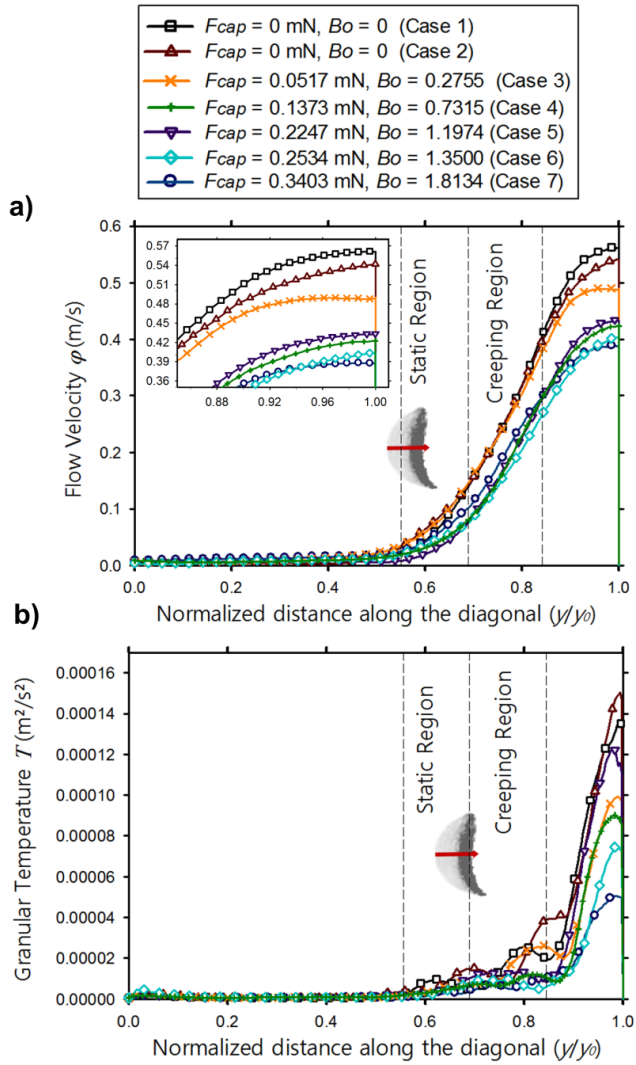

Figure 12. a) Average flow velocity $\varphi$ profile along the diagonal $(y/y_0)$ normal to the free surface. b) Average granular temperature profile along the diagonal $(y/y_0)$ normal to the free surface. Particles size $r = 1.25$ mm and drum rotation speed $\omega = 25$ rpm.

Figure 12 (a) and (b) show the spatial distribution along the $y$ axis of the averaged flow velocity $\varphi$ and the granular temperature $T$ respectively, as obtained using equation (10). The granular temperature and the flow velocity are plotted from the bottom of the bed, traversing normal to the bed surface (i.e. along axis $y$) up to the top of the bed. The distance along $y$ is normalized by the width of the bed that goes from the bottom of the bed to its top (i.e. $y/y_0$).

In Figure 12 (a), as $y/y_0$ increases, right after the static region, the flow velocity shows a rapid rise followed by a steady increase over the remaining range, and finally reaches maxima. This flow velocity maxima decreases with the capillary force with the highest maxima observed for dry glass beads. We attribute this decrease to the formation of clusters network. As the clusters increase in size, the bed starts to slip as a bulk of particles interconnected by capillary forces, where each layer of particles in the bulk is slowed down by the cohesive forces excreted by the layer below it. The capillary force between the particles leads to a higher drag force in the opposite direction of the flow. Higher drag force slows down particle movement along the creeping and the flowing regions. Similar observation was obtained by Liao et al. [55] upon increasing the viscosity of the interstitial liquid. By adding glycerin rather than water, these authors found that the flowing layer becomes thicker and the flow velocity decreases.

Figure 12 (b) shows the granular temperature profile along the $y$ axis perpendicular to the flow. Near the wall, because of the solid-like movement of the particles, the granular temperature is almost absent. In the static region, the granular temperature starts to increase slowly, then increases dramatically and reaches a peak in the flowing region. This indicates a solid-like displacement in the static region and high collision of particles in the flowing regions, confirming the results obtained in Figure 11. We notice that cases 1 and 2 have the highest granular temperature maxima, and case 7 have the lowest one. However, the spots of the granular temperature are not uniform and are dispersed in the creeping and flowing regions (see Figure 11). This means that the magnitudes of the peaks along the diagonal of the curves in Figure 12 (b) are not representative of the effect of the capillary forces on the granular temperature. For this reason, we will examine next the flow characteristics along the surface of the bed.

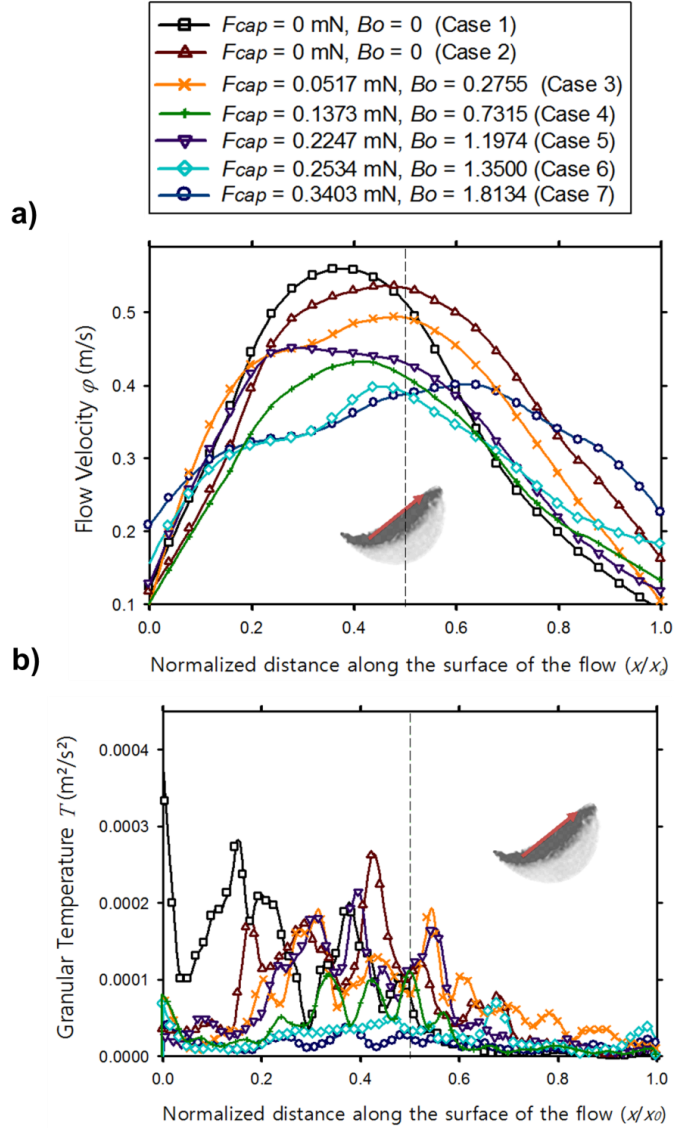

Figure 13. a) Average flow velocity $\varphi$ profile along the surface of the flow $(x/x_0)$, b) Average granular temperature profile along the surface of the flow $(x/x_0)$. Particles size $r = 1.25$ mm and drum rotation speed $\omega = 25$ rpm.

Similarly to Figure 12, we report in Figure 13 (a) and (b) the flow velocity profile and the granular temperature of the particles respectively, along the surface of the flow. They are both plotted at the surface of the flowing layer, starting from the left side of the drum, crossing the midpoint of the flow to the right side of the drum. The distance along the $x$ axis is normalized by the length of the surface outline (i.e. $x/x_0$), and the width of the averaged zone $L$ here is also equal to 4×$r$. In Figure 13 (a), as expected, the maximum of the flow velocity profile decreases with the capillary force. Moreover, the surface profiles exhibit a parabolic shape, skewed to left for the dry case and start to shift to the right as the capillary force increases. Boateng and Barr [56] observed the same skewness of surface flow in the case of dry materials. They attributed it to the ability of the materials to dissipate energy through collision. For the dry case, particles go faster past the mid-chord position, but, once mixed with liquid, they reach their maxima at the midpoint and a symmetric profile is obtained, indicating that the particles flux into the flowing layer equals the one leaving it. The shifting to the right can also be observed in the case of the averaged granular temperature shown in Figure 13 (b), where the intensity of the granular temperature is the highest on the left side of the drum. This is mainly because the velocity of the particles falling from the right side of the bed and colliding on the left side is reduced by the capillary cohesion. Furthermore, the peaks of the granular temperature decrease with the capillary force. Once again, it can be inferred here that the kinetic energy dissipation in the surface flow due to collision is reduced in favor of energy dissipation due to capillary forces.

Let's examine the effect of the capillary force further by investigating the shear stress in the flowing region. Since the velocity profile in the midpoint of the flowing region is linear (see Figure 12 (a)), we can compute the shear stress on this region using the following formula:

$$\dot{\delta} = \frac{\Delta v_{av}}{\Delta y}. \qquad\qquad Eq.\ (11)$$

In Figure 14, we present the shear stress $\dot{\delta}$ in the flowing region of the drum for particle size 1.25 mm, as a function of capillary force and rotation speed. We vary the rotation speed for wet particles with 100% ($V/V$) water (case 7) and we present its curve with cross red symbols in Figure 14. We also vary the capillary force for drum rotation speed of 25 rpm and we present its curve with a square symbols in Figure 14. The values of the shear stress were calculated from the slope of the linear part of the velocity profiles using Equation (11). The shear stress increases linearly with an increase of the rotation speed. This is a direct result of the increase of the flow velocity. An increase of the flow velocity increases the sliding of the flowing layer, which in turn increases the shear stress in that region. However, the shear stress slightly decreases with the increase of the capillary force, due to higher kinetic energy dissipation and less particle collisions. The drag force exerted by the core of the flowing region on the particles on the free surface increases with the capillary force (i.e. the deeper region with relatively slower speed decelerate the upper flow by means of capillary forces). This reduces the flow velocity, and thus decreases the shear stress. Overall, the effect the capillary force on the flow velocity on the surface and the shear stress is low comparing to that of the rotation speed of the drum.

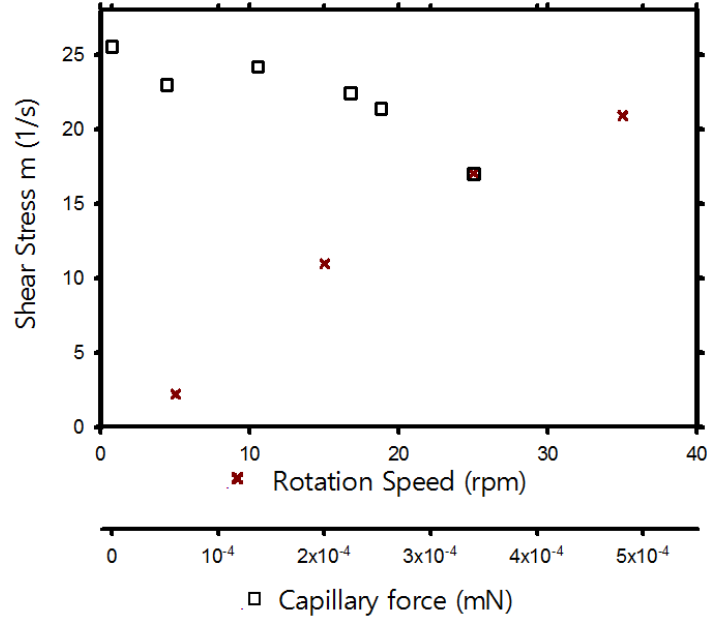

Figure 14. Shear stress in the flowing region as a function of the capillary force at 25 rpm, and as a function of rotation speed at 100% (v/v) water, case 7.

## 3.3 Scaling of wet granular flow

The concept of similarity based on dimensional analysis states that two processes might be considered similar if all involved length scales are proportional, and all dimensionless numbers needed to describe them have the same value [57, 16].

Five dimensionless groups are presented in equation (12) including Froude number $F_r$ (ratio of inertial force to gravity force), granular Weber number $We$ (i.e. ratio of inertial forces to capillary forces), particle volume to liquid ratio, capillary number $C_a$, and drum fill ratio $\beta$.

$$F_r = \sqrt{\frac{(D-2r)w^2}{2g}}, \quad We = \frac{\rho_p r v^2}{\gamma \cos\theta}, \quad \frac{N_p r^3}{v_{liq}},$$

$$Ca = \frac{v_r \mu}{\gamma \cos\theta}, \quad \beta,$$

*Eq. (12)*

where $N_p$ is the total number of particles. Notice that the Weber and Capillary numbers in equation (12) take into consideration the effect of the particle-liquid contact angle $\theta$.

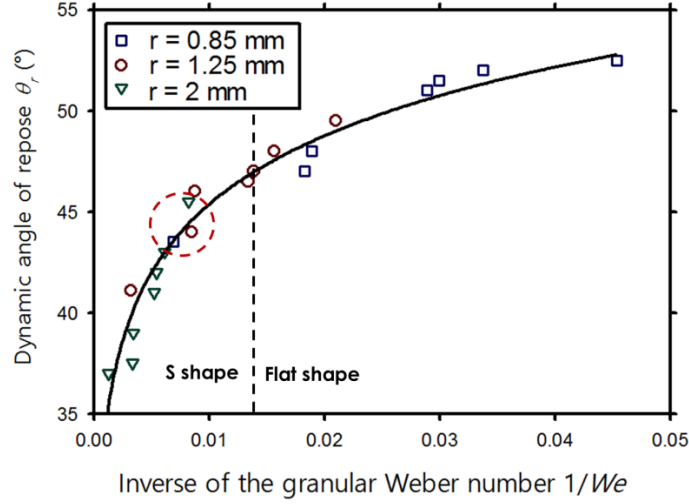

Figure 15. Dynamic angle of repose $\theta_r$ with its respective fitted curve, plotted against the granular Weber number. Drum rotation speed: 25 rpm.

In our study, we have kept the Froude number, liquid-particle volume ratio, and drum fill percentage constant among all experiments. We scale the dynamic angle of repose of wet particles using the Weber number. As empathized previously, the effect of capillary force on the flow velocity is low comparing to the effect of the rotation speed of the drum. Hence, we will define the average velocity in the Weber number equation equal to the rotation speed of the drum times its radius (i.e. $w \times D/2$), and we focus on the effect of the Weber number on the flow where the capillary forces are dominant (i.e. capillary number $C_a \ll 1$). In Figure 15, we show the evolution of the dynamic angle of repose as a function of the inverse of the granular Weber number. We vary the Weber number by varying the particle size for different capillary forces (cases 3 to 7 in Table 1, for 0.85, 1.25 and 2 mm particles) for a fixed rotation speed equal to 25 rpm. The dynamic angle of repose increases with the inverse of the Weber number and shows a logarithmic trend with the following equation of the fit:

$$\theta_r = 68.03 + 4.91 \ln(1/We).  \qquad Eq.\ (13)$$

Points with close Weber number are expected to have close dynamic angle of repose. This allows to maintain dynamic similarity after scaling the particle size. We will test this hypothesis by scaling the particle size and keeping the Weber number constant by increasing the capillary force accordingly.

The points inside the red circle in Figure 15 have different particle size: 2, 1.25 and 0.85 mm and three different cases: case 7, case 4 and case 3 respectively (case 7: Water with non-silanized glass beads, case 4: mixture of 10% ethanol-90% water ($V/V$) with silanized glass beads, case 3: mixture of 20% ethanol-80% water ($V/V$) with silanized glass beads, see Table 1), but they have the same

Weber number. We compare those three points and we check whether they have the same dynamic angle of repose for different rotation speed of the drum.

Figure 16 presents the dynamic angle of repose of wet samples for different rotation speeds. While the red symbols represent the three points that we chose (inside the red circle in Figure 15) of different particle sizes with almost the same Weber number, solid blue symbols represent non-silanized particles of size 0.85 mm mixed with 100% (V/V) water (case 7) and lower Weber number.

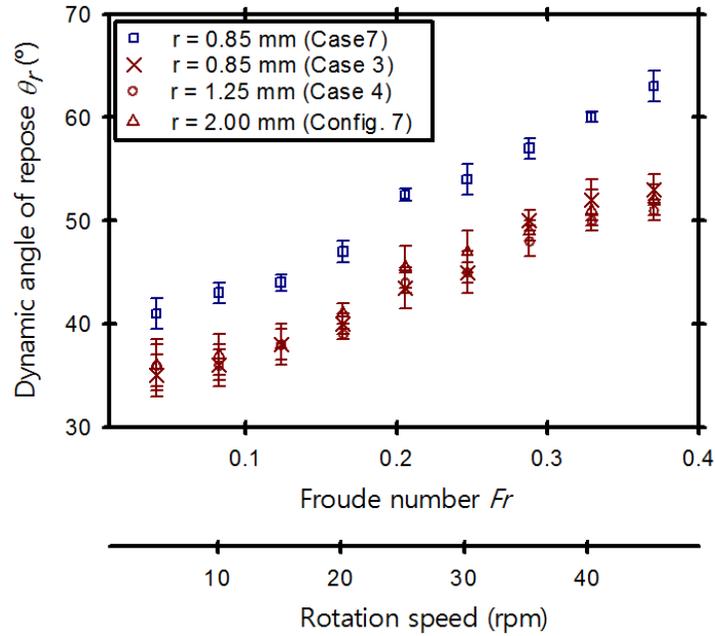

Figure 16. Dynamic angle of repose as a function of the rotation speed for different particle sizes in the wet case.

The dynamic angle of repose of the chosen points (i.e. particle sizes 2, 1.25 and 0.85 mm) fall onto a single curve, indicating that 2 mm particles can be rescaled to 0.85 mm or 1.25 mm sizes if the Weber number is kept constant. This means that, experimentally, the scaling approach works in the considered flow regimes (rolling and cascading).

Let us explore further this scaling and look at the flow velocity $\varphi$ of these three points (inside the red circle in Figure 15) with the same Weber number. Similarly to subsection 3.2, we plot in Figure 17 (a) and (b), the flow velocity profiles at the midpoint of the flow along the diagonal and on the surface of the flow respectively. The blue square symbols here also represent wet particles with 100% (V/V) water (case 7, $r = 0.85$ mm). The profiles of the red symbols curves are close to one another comparing to the blue curve, especially in the case of the diagonal $\varphi$ profile in Figure 17 (a). The flow velocity at the midpoint of the flowing region is almost the same for the three particle sizes (in red). In Figure 17 (b), small lateral variations are observed for flow velocity along the surface of the red curves, especially in the case of 2 mm particle size, but still, here also, they show similar profiles comparing to the blue case.

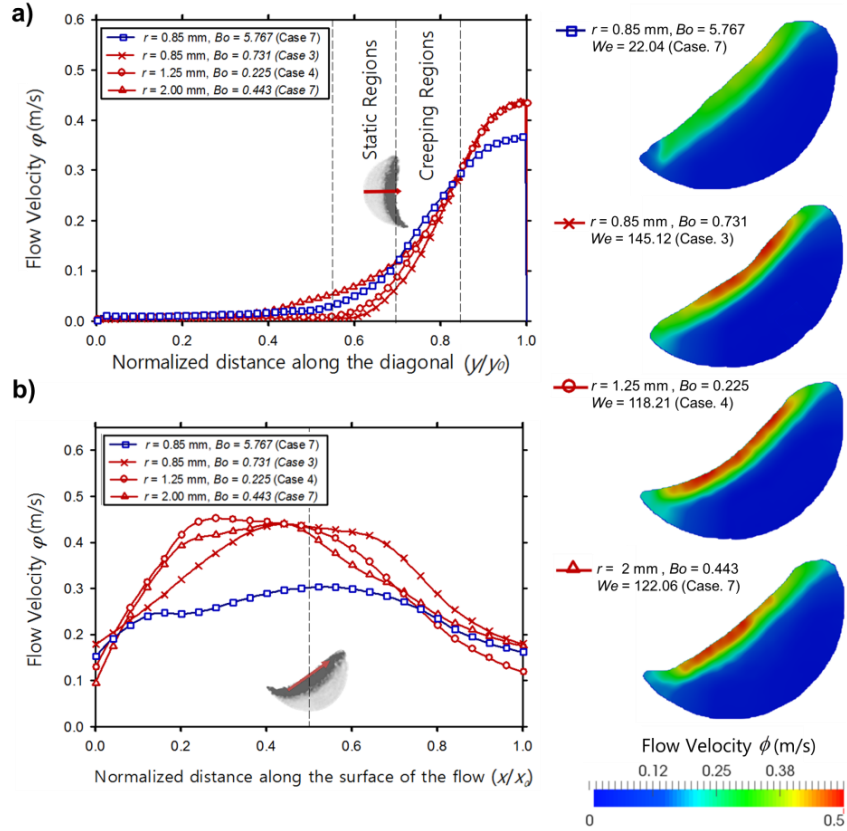

Figure 17. a) Average flow velocity $\varphi$ profile along the diagonal $(y/y_0)$ normal to the free surface, b) Average flow velocity $\varphi$ profile along the surface of the flow $(x/x_0)$. Drum rotation speed $\omega = 25$ rpm.

Since the dynamic angle of repose of dry particles can be scaled with the Froude number as we already showed in Figure 7, we will develop a scaling equation that take into consideration the rotation speed. We will focus on the ratio of the angle of slope of the wet over that of the dry case. In Figure 18, we plot $\theta_{r,\,wet} / \theta_{r,\,dry}$ as a function of the inverse of the Weber number for different rotation speeds and different particle sizes, where $\theta_{r,\,wet}$ is the dynamic angle of repose in the wet case, and $\theta_{r,\,dry}$ is the dynamic angle of repose in the dry case. $\theta_{r,\,wet} / \theta_{r,\,dry}$ increases linearly with the inverse of the Weber number and all the points seem to collapse into a single master curve with a standard error of 0.0303. Figure 18 can be used to predict and control the flow of the particles from the surface properties of glass beads and the adequate ethanol-water fraction.

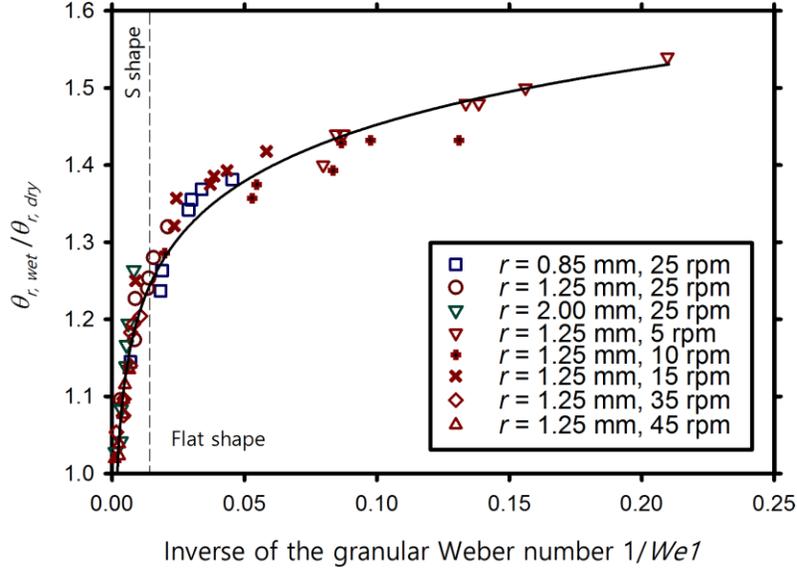

Figure 18. Ratio of the wet over the dry dynamic angle of repose, $\theta_{r,\,wet} / \theta_{r,\,dry}$ plotted against the granular Weber number with its fitted curve.

The curve fit obtained using $We$ from the data in Figure 18 is described by the following equation:

$$\theta_{r,\,wet} / \theta_{r,\,dry} = 1.64 + 0.1048\ln(1/We) \qquad Eq.\ (14)$$

By incorporating the Froude number (see Figure 7), we obtain:

$$\theta_{r,\,wet} = 36.408 + 139.4Fr + 2.33\ln(1/We) + 8.908Fr\ln(1/We) \qquad Eq.\ (15)$$

## 4. Conclusion

Experiments on the continuous flow of granular material in a rotating drum were reported in this paper. The effect of the capillary forces and the particle size on the flow of the bed was investigated for different rotation speeds. We have extracted granular temperature and flow velocity profiles by means of statistical analysis and PIV. We established a scaling relationship using the Froude and the Weber number and we showed that the dynamic angle of repose and the shape of the free surface can be varied and controlled by a combination of rotation rate of the drum and capillary forces. At the end of the paper, we established a simple graphical method that correlates the liquid composition and glass beads properties to the dynamic angle of repose.

We demonstrated that chemical salinization of the glass beads allows to alter the capillary forces between the particles, allowing the investigation of the flow of granular assembly over a wide range of capillary forces. We showed that the presence of liquid increases the depth of the flowing layer and reduces the mobility of the particles. Granular temperature snapshots showed that capillary forces reduce particles collisions in favor of kinetic energy dissipation due to cohesion forces. When the cohesion between particles increases, particles become closely packed and act as clusters rather than individually. Capillary force leads to greater drag force exerted by the core of the bed on the flowing region, resulting in a slower flow velocity and thus, decreases the shear stress in the flowing region. Then, we proposed a scaling methodology that ensures similarity of the dynamic angle of repose by keeping the Weber number constant after scaling the particle radius. We were able to obtain similar bed flow for different particle sizes, confirming that the proposed scaling approach works for the rolling and cascading regimes.

The results reported here open several prospects. For instance, it would be interesting to include the effect of the liquid viscosity on the rather dynamic bed flow which will be the focus of future work. It would be of great interest also to consider how well the scaling methodology works in other flowing regimes such as the cataracting regime. Moreover, this paper provided some important clarifications regarding the capillary force effects on the flow of dynamic granular assemblies that can be applied to other granular flow (e.g. Silo, inclined plane, chute flow, ect.).

# Acknowledgements

We thank Andries van Swaaij (Dries), Dr. Marco Ramaioli and Bert Scheper for their help. We also thank Prof. Johannes Khinast for his feedbacks. Financial support through the "T-MAPPP" project of the European-Union-Funded Marie Curie Initial Training Network FP7 (ITN607453) is also acknowledged.

# List of symbols

| Symbol | Description | Units |
|---|---|---|
| $Ca$ | Capillary number | [-] |
| $D$ | Drum diameter | [m] |
| $d_0$ | Distance between 2 particles | [m] |
| $F_{cap}$ | Capillary force | [N] |
| $Fr$ | Froude number | [-] |
| $g$ | Gravitational acceleration | [m/s$^2$] |
| $L$ | Width of the drum | [m] |
| $N_p$ | Total number of particles in the drum | [-] |
| $r$ | Particle radius | [m] |
| $Ti$ | Granular Temperature | [m²/s²] |
| $V_{bed}$ | Volume of the bed | [m³] |
| $V_{liq}$ | Volumetric liquid content | [m³] |
| $v$ | Particle velocity | [m/s] |
| $v_r$ | Relative velocity | [m/s] |
| $v_\Omega$ | Angular velocity | [m/s] |
| $We$ | Weber number | [-] |
| $s$ | Granule saturation | [-] |
| $Y$ | Young's modulus Y | [N/m²] |
| $\nu$ | Poisson ratio ν | [-] |
| $\theta$ | Contact angle | [°] |
| $\theta r$ | Dynamic angle of repose | [°] |
| $\theta m$ | Maximum angle of stability | [°] |
| $\theta s$ | Lower dynamic angle of repose | [°] |
| $\gamma$ | Surface tension | [N/m] |
| $\omega$ | Rotation speed | [s$^{-1}$] |
| $\varrho_p$ | Particle density | [Kg/m³] |
| $\alpha$ | Coarse-graining width | [m] |
| $\beta$ | Filling level of the drum | [-] |
| $\mu$ | Viscosity | [Pa.s] |
| $\varphi$ | Flow velocity | [m/s] |
| $\dot{\delta}$ | Shear stress | [1/s] |

# References

[1] H. Henein, J. K. Brimacombe, and A.P. Watkinson, Experimental study of transverse bed motion in rotary kilns, Metall. Trans. B 14 (1983), 191–205.

[2] J. Rajchenbach, Flow in powders: From discrete avalanches to continuous regime, Phys. Rev. Lett. 65, (1990), 2221.

[3] N. Taberlet, P. Richard, E.J. Hinch, The S-shape of a granular pile in a rotating drum, Phys. Rev. E 73 (2006), 050301.

[4] A.V. Orpe, D.V. Khakhar, Scaling relations for granular flow in quasi two‐dimensional rotating cylinders, Phys. Rev. E 64 (2001), 031302.

[5] T. Elperin and A. Vikhansky, Granular flow in a rotating cylindrical drum, Europhys. Lett. 42 (1998), 619-23.

[6] L.T. Sheng, W.C. Chang, and S.S. Hsiau, Influence of particle surface roughness on creeping granular motion, Phys. Rev. E 94 (2016), 012903.

[7] M. Nakagawa, S. Altobelli, A. Caprihan, E. Fukushima, and E. Jeong, Non-invasive measurements of granular flows by magnetic resonance imaging, Exp. Fluids 16 (1993), 54-60.

[8] Komatsu, T., S. Inagaki, N. Nakagawa, & S. Nasuno.Creep Motion in a Granular Pile Exhibiting Steady Surface Flow. Phys. Rev. lett. 86 (2001), 101103.

[9] N. Jain, J. M. Ottino and R. M. Lueptow, An experimental study of the flowing granular layer in a rotating tumbler. Phys. Fluids 14 (2002), 572–582.

[10] P. Tegzes, T. Vicsek, P. Schiffer, Development of correlations in the dynamics of wet granular avalanches, Phys. Rev. E 67 (2003), 051303.

[11] H. Schubert, Capillary forces - modeling and application in particulate technology, Powder Tech. 37 (1984), 705 – 176.

[12] R. Brewster, G. S. Grest, and A. J. Levine, Effects of cohesion on the surface angle and velocity profiles of granular material in a rotating drum, Phys. Rev. E 79 (2009), 011305, 1-7.

[13] S.H. Chou, C.C. Liao, S.S. Hsiau, An experimental study on the effect of liquid content and viscosity on particle segregation in a rotating drum. Powder Tech. 201 (2010), 266–272.

[14] P. Y. Liu, R. Y. Yanga, and A. B. Yu Dynamics of wet particles in rotating drums: Effect of liquid surface tension, Phys. Fluids 23 (2011), 013304, 1-9.

[15] A. Samadani, A. Kudrolli, Segregation transitions in wet granular matter, Phys. Rev. Lett. 85 (2000), 5102–5105.

[16] A. Jarray, V. Magnanimo, M. Ramaioli, and S. Luding. Scaling of wet granular flows in a rotating drum, EPJ Web of Conferences, Powders and Grains 140 (2017), 03078, 1-4.

[17] S. Nowak, A. Samadani, and A. Kudrolli. Maximum angle of stability of a wet granular pile, Nat. Phys. 1 (2005), 50-52.

[18] Q. Xu, A.V. Orpe, and A. Kudrolli, Lubrication effects on the flow of wet granular materials, Phys. Rev. E 76 (2007), 031302.

[19] C. Soria-Hoyo, J.M. Valverde, A. Castellanos. Avalanches in moistened beds of glass beads. Powder Technology 196 (2009), 257–262.

[20] D. Hornbaker, R. Albert, I. Albert, A.L. Barabasi, P. Schiffer, What keeps sandcastles standing?, Nature 387 (1997), 765-765.

[21] M. M. Kohonena, Geromichalosb D., Scheelb M., Schierb C., S. Herminghaus. On capillary bridges in wet granular materials, Physica A 339, 1–2 (2004), 7–15.

[22] J. E. Fiscina, G. Lumay, F. Ludewig, and N. Vandewalle, Compaction Dynamics of Wet Granular Assemblies, Phys. Rev. lett. 105 (2010), 048001.

[23] R. Schwarze, A. Gladkyy, F. Uhlig, S. Luding, Rheology of weakly wetted granular materials - a comparison of experimental and numerical data, Granular Matter 15, 4 (2013), 455-465.

[24] Z. Fournier, D. Geromichalos, S. Herminghaus et al., Mechanical properties of wet granular materials J. Phys. Condens. Matter. 17 (2005), S477–S502.

[25] T. Gillespie and W. J. Settineri, The effect of capillary liquid on the force of adhesion between spherical solid particles, J. Colloid Interface Sci. 24 (1967), 199-202.

[26] R. A. Fisher. On the capillary forces in an ideal soil, J. Agric. Sci. 16 (1926), 492-505.

[27] T. C. Halsey and A. J. Levine, How Sandcastles Fall, Phys. Rev. Lett. 80 (1998), 3141.

[28] J.N Israelachvili, Intermolecular and Surface Forces (Academic Press), 2010, University of California.

[29] Y.I. Rabinovich, M.S. Esayanur and B.M. Moudgil, Capillary forces between two spheres with a fixed volume liquid bridge: theory and experiment, Langmuir 21 (2005), 10992-7.

[30] P. S. Raux, H. Cockenpot, M. Ramaioli, D. Quéré and C. Clanet, Wicking in a Powder, Langmuir 29 (11) (2013), 3636–3644.

[31] A.F. Stalder, T. Melchior, M. Müller, D. Sage, T. Blu, Low-Bond Axisymmetric Drop Shape Analysis for Surface Tension and Contact Angle Measurements of Sessile Drops, Colloids Surf. A 364 (2010), 72-81.

[32] G. Vazquez, E. Alvarez, J. M. Navaza, Surface Tension of Alcohol Water + Water from 20 to 50 .degree.C, J. Chem. Eng. Data 40 (3) (1995), 611–614.

[33] M. H. Klein Schaarsberg, I.R. Peters, M. Stern, K. Dodge et al., From splashing to bouncing: The influence of viscosity on the impact of suspension droplets on a solid surface, Phys. Rev. E 93 (2016) 062609.

[34] Q. Xu, I. Peters, S. Wilken, E. Brown, and H. Jaeger, Fast Imaging Technique to study Drop Impact Dynamics of non-Newtonian Fluids, J. Vis. Exp. 85, 51249 (2014).

[35] M. Tagawa, K. J. Gotoh, Y. Nakagawa, Penetration of Water/Ethanol Mixtures into Silanized Silica Fibrous Assemblies, J. Adhesion Sci. Technol. 12 (1998), 1341-1353.

[36] N. Jain, J. M. Ottino, and R. M. Lueptow, Regimes of segregation and mixing in combined size and density granular systems: an experimental study, Granul. Matter 7 (2005), 69-81.

[37] G. Juarez, P. Chen and R. M. Lueptow, Transition to centrifuging granular flow in rotating tumblers: a modified Froude number, New J. Phys. 13 (2011), 053055.

[38] J. Litster and B. Ennis, Springer, The Science and Engineering of Granulation Processes. volume 15, Netherlands, 250.

[39] A. D. Salman, M.J Hounslow and J.P.K Seville, ELSEVIER, Handbook of Powder Technology, Granulation, Volume 11, 2007.

[40] J.-Y. Tinevez, N. Perry et al. TrackMate: An open and extensible platform for single-particle tracking, Methods 115 (2017), 80–90.

[41]. G. Lowe, Distinctive image features from scale-invariant keypoints. Int. J. Comput. Vision (2004) 60, 91–110.

[42] A. Henderson, A ParaView Guide: Parallel Visualization Application, Kitware Inc. (2007).

[43] A. L. Chaua, X. Lia, W. Yu, Convex and concave hulls for classification with support vector machine, Neurocomputing 122 (2013), 198–209.

[45] W. Thielicke and E.J. Stamhuis, PIVlab – Towards User-friendly, Affordable and Accurate Digital Particle Image Velocimetry in MATLAB. J. Open Res. Softw. 2 (1) (2014), e30, 1-10.

[46] H. Abram Clark, L. Kondic, and R. P. Behringer. Steady flow dynamics during granular impact, Phys. Rev. E 93 (2016), 050901.

[47] A. Seguin, Y. Bertho, F. Martinez, J. Crassous, and P. Gondret,Experimental velocity fields and forces for a cylinder penetrating into a granular medium, Phys. Rev. E 87 (2013), 012201.

[48] T. Weinhart, C. Labra, S.Luding, J. Y. Ooi, Influence of coarse-graining parameters on the analysis of DEM simulations of silo flow, Powder Tech. 293 (2016) 138–148.

[49] T., Weinhart, R. Hartkamp, A. R. Thornton, and S. Luding. Coarse-grained local and objective continuum description of three-dimensional granular flows down an inclined surface. Phys. Fluids 25 (7) (2013), 1-34.

[50] Y. Zhang and C. S. Campbell, The interface between fluid-like and solid-like behaviour intwo-dimensional granular flows. J. Fluid. Mech. 237 (1992), 541-568.

[51] H. Ahn, C. E. Brennen and R. H. Sabersky, Measurements of velocity, velocity fluctuation, density, and stresses in chute flows of granular materials, Trans. ASME J. Appl. Mech. 58 (1991), 792-803.

[52] D. Bonamy, P.-H Chavanis, P.-P. Cortet, F. Daviaud, B. Dubrulle, M. Renouf, Euler-like modelling of dense granular flows: application to a rotating drum, Euro. Phys. J. B 68 (4) (2009), 619-627.

[53] H.T. Chou and C.F. Lee, Cross-sectional and axial flow characteristics of dry granular material in rotating drums, Granular Matter 11 (2009), 13–32.

[54] D. Bonamy, F. Daviaud, L. Laurent, M. Bonetti, J.-P. Bouchaud, Multiscale Clustering in Granular Surface Flows, Phys. Rev. Lett. 89 (2002), 034301.

[55] C. C. Liao, S. S. Hsiau and Kiwing To. Granular dynamics of a slurry in a rotating drum. Phys. Rev. E 82 (2010), 010302R.

[56] A. A. Boateng and P. V. Barr. Granular flow behaviour in the transverse plane of a partially filled rotating cylinder. J. Fluid Mech. 330 (1997), 233-249.

[57] F.M. Henderson, Open channel flow. Macmillan Publishing Co., Inc., (New York, 1966).

# Appendix A

Figure A.1 shows the maximum angle of stability $\theta_m$ as a function of the Froude number $Fr$. $\theta_m$ increases slowly with the Froude number and exhibit high variability especially at low rotation speeds. When the Froude number increases, the rotation time becomes too short compared to the avalanche duration; consequently, $\theta_m - \theta_r$ decreases, which explain the decrease of the variability with the Froude number.

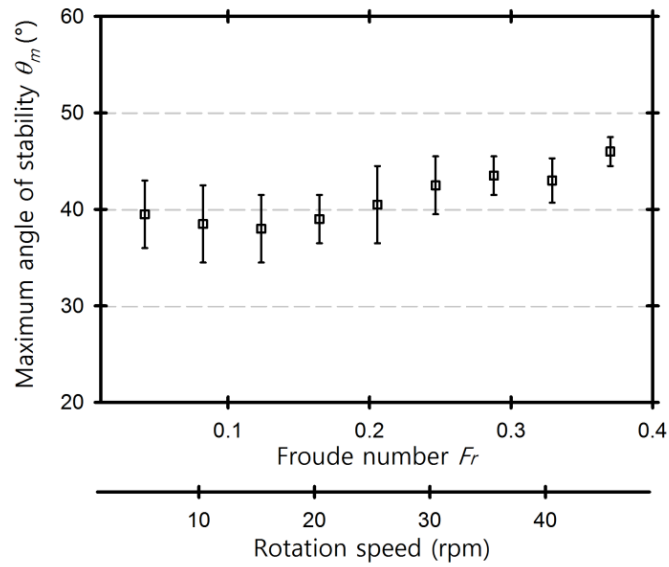

Figure A.1. Maximum angle of stability $\theta_m$ as a function of the rotation speed and the Froude number $Fr$, for the dry case, for particle size $r = 1.25$ mm.

# Appendix B

Figure B.1 shows the velocity $v$ components of the bed in a rotating drum for different capillary forces. The three zones; flowing, creeping and static are easily distinguishable Figure B.1 (a). The wall driven (i.e. solid body rotating with the drum) velocity is observed at the head of the jellyfish at the bottom-left and the flowing zone is spread on the tail of the jellyfish shape. As the capillary force increases the tail of the jellyfish shrinks indicating denser distribution of the velocity components and more organized flow, confirming the results shown in Figure 10. The median angle made by the cloud of the velocity components in the creeping region is approximately equal to $(\theta r + \theta s)/2$. The particles as well as the velocity components distribution are less scattered and denser in this region comparing to the other regions (see Figure B.1 (a)). As explained by Komatsu [8], the particles in the static region inhibit the movement of those outside this region. As a result, the flow in the creeping region experience high resistance and the creep motion is induced right under the flowing region.

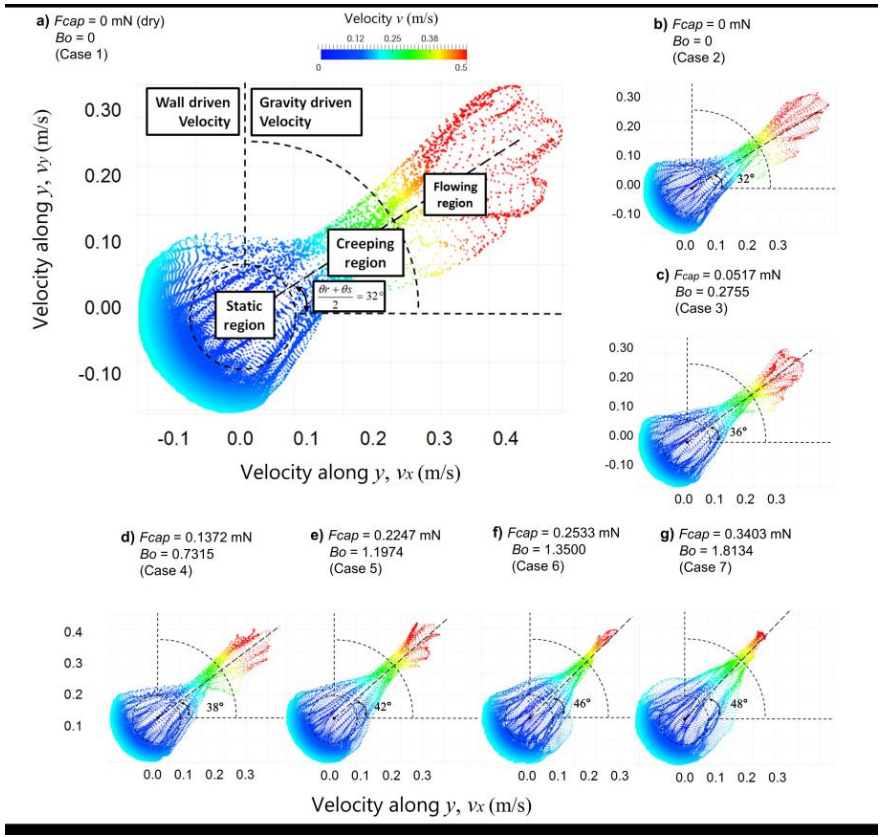

Figure B.1. Snapshots of the velocity's vertical and horizontal components in the rotating drum for different capillary forces. Particles size $r = 1.25$ mm and drum rotation speed $\omega = 25$ rpm.